\documentclass[aps,prb,twocolumn,showpacs,superscriptaddress,longbibliography]{revtex4-2}
\usepackage{graphicx}
\usepackage{amsfonts,amssymb,amsmath}
\usepackage{bm}
\usepackage{xcolor}
\begin{document}
\title{Quantum computing of magnetic-skyrmion-like patterns in 
Heisenberg ferromagnets}

\date{\today}
 
\author{Matej Komelj}
\email{matej.komelj@ijs.si}
\affiliation{Jo\v zef Stefan Institute, Jamova cesta 39, SI-1000 Ljubljana,
  Slovenia}
%\affiliation{\color{black} Faculty of Mathematics and Physics, University of 
%Ljubljana, 1000 Ljubljana, Slovenia}
\author{Vinko Sr\v san}
\affiliation{Jo\v zef Stefan Institute, Jamova cesta 39, SI-1000 Ljubljana,
  Slovenia}
\affiliation{International Postgraduate School Jo\v zef Stefan, Jamova cesta 39,
SI-1000 Ljubljana, Slovenia}
\author{Kristina \v Zu\v zek}
\affiliation{Jo\v zef Stefan Institute, Jamova cesta 39, SI-1000 Ljubljana,
  Slovenia}
\affiliation{International Postgraduate School Jo\v zef Stefan, Jamova cesta 39,
SI-1000 Ljubljana, Slovenia}
\author{Sa\v so \v Sturm}
\affiliation{Jo\v zef Stefan Institute, Jamova cesta 39, SI-1000 Ljubljana,
  Slovenia}
\affiliation{International Postgraduate School Jo\v zef Stefan, Jamova cesta 39,
SI-1000 Ljubljana, Slovenia}
\begin{abstract}
We diagonalize the quantum two-dimensional spin-1/2 Heisenberg model 
with Dzyaloshinskii-Moriya interaction (DMI) by applying the variational
quantum eigensolver, running on a quantum-computer simulator, {\color{black}
which turns out to be a more efficient approach than a classical direct
diagonalization for systems with more than 17 sites.}  The calculated 
external-magnetic-field dependence of 
the total energy, of the magnetization, as well as of the topological charge 
exhibits a distinctive 
discontinuity which hints for the existence of zero-temperature magnetic 
{\color{black} skyrmions-like structures} at the quantum level, controlled by 
the combination of the 
exchange-coupling and  the DMI parameters. The potentially measurable jump in 
the magnetization upon changing the field indicates the investigated objects as 
stable enough for eventual applications in spintronics or even as information
carriers.
\end{abstract}
\maketitle
The exciting field of quantum computing is interesting from the point
of view of developing the required hardware and software, as well as by means
of applying the respective algorithms for certain tasks, where they might at 
least outperform classical approaches or even make unsolved problems 
solvable\cite{ref1}. Although the number of commercially-available 
platforms for quantum 
computations constantly grows,
the implementation of the basic components, 
for example, of the information carriers in terms of qubits and of the 
respective
logical gates is not yet standardized\cite{PRXQuantum.2.017001}. 
Any attempts of simplifying the 
existing set-ups, particularly by avoiding the necessity for operating
at extremely low temperatures, are highly desired. In this manner, 
magnetic skyrmions might represent a considerable alternative, worth of 
exploring\cite{PhysRevLett.127.067201,PhysRevLett.130.106701}.  \par
Vortex-like magnetization $\vec{m}(x,y)$ patterns, here confined to the 
$x$-$y$ plane, are characterized by the topological charge $Q$ (sometimes
referred as the topological number or index)\cite{PhysRevB.110.094442}:
\begin{equation}
{\color{black}Q}={1\over 4\pi}\int\vec{m}(x,y)\cdot\left({\partial\vec{m}(x,y)\over\partial x}
\times{\partial\vec{m}(x,y)\over\partial y}\right)dxdy.
\end{equation}
Strictly speaking, a proper skyrmion is distinguished by an integer $|Q|\ge1$
and its stability arises from a topological protection. However, according
to a less rigid definition, it is an energy barrier which stabilizes the 
corresponding
magnetization textures, which creation and annihilation are triggered by 
external parameters, such as surrounding magnetic 
fields\cite{PhysRevB.93.174403}. The stability, induced
either by a topological protection or by an energy barrier, 
and the ability to switch them on/off  give skyrmions the potential to 
be applied in future memory devices.\par
Whereas macroscopic skyrmions exist at certain finite-temperature ranges, 
a theory predicts them at quantum level for 
$T=0$\cite{PhysRevB.92.245436,PhysRevResearch.4.043113}. However, even at that case
a direct diagonalization of lattice
quantum models with magnetic interactions is demanding since the set of 
possible states 
grows exponentially with the number of sites, representing single magnetic 
moments\cite{PhysRevResearch.4.023111}.
The basic Hamiltonian, which solutions might include
a skyrmionic phase in dependence of the parameters, is associated with
the two-dimensional $1/2$-spin anisotropic Heisenberg (XXZ) model in a
perpendicular external magnetic field $B_z$:
\begin{eqnarray}
H=J_\parallel\sum_{<i,j>}\left(\sigma^x_i\sigma^x_j+\sigma^y_i\sigma^y_j\right)+
J_\perp\sum_{<i,j>} \sigma^z_i\sigma^z_j+\\
+\sum_{<i,j>}\vec{D}_{ij}\cdot\left(\vec{\sigma}_i\times\vec{\sigma}_j\right)+B_z\sum_i
\sigma^z_i, \nonumber
\end{eqnarray}
where $\vec{\sigma}_i=(\sigma_i^x,\sigma_i^y,\sigma_i^z)$ are Pauli matrices
and $<i,j>$ denotes a sum over the nearest neighbors. 
The exchange coupling constants $J_\parallel<0$ and $J_\perp>0$ imply
a ferromagnetic ordering with the $x$-$y$ easy-plane anisotropy. 
Note, the results
and conclusions presented below are qualitatively the same if
an isotropic exchange $J_\parallel=J_\perp$ is considered, but in that case
the $z$-contribution to the magnetization is the dominant, hence the in-plane
vortex patterns are less pronounced. The third term, describing the 
antisymmetric exchange in a form of the Dzyaloshinskii–Moriya interaction 
(DMI)\cite{DZYALOSHINSKY1958241,1960PhRv..120...91M} , is the key for a 
formation of skyrmion-like patterns. It favors 
neighboring magnetic moments at the sites $i$ and $j$ being perpendicular to 
each other,
either parallel or anti-parallel to the coupling vector $\vec{D}_{ij}$. 
However, the presence of the DMI is not the only way to stabilize skyrmion-like
structures on quantum level. For example, frustrations in a Heisenberg 
model with the next-nearest-neighbors exchange interactions exhibit similar
solutions\cite{PhysRevX.9.041063} but these are not a subject of the present
paper.
An application of a Hamiltonian like (2) on a multiple-qubits
state is straight forward since it is formulated by Pauli matrices, which
are among the basic logical gates in quantum computing. Therefore it makes
sense to explore its properties by diagonalizing it using the 
variational quantum eigensolver (VQE)\cite{Peruzzo2014}, which is a common
tool for finding the ground state of a physical system in quantum chemistry 
and quantum magnetism\cite{2022PhR...986....1T}. {\color{black} A migration
to an appropriate hardware should be feasible.} \par
For the sake of simplicity we fix $|J_\parallel|$=$|\vec{D}_{ij}|$ as 
the unit and impose a ferromagnetic in-plane ordering by setting 
$J_\parallel<0$.
The DMI vectors $\vec{D}_{ij}$ with two different directions are considered:
either parallel or perpendicular to the vector $\vec{R}_{ij}$ 
pointing from the site $i$ to the site $j$, as sketched in FIG. 1. 
\begin{figure}
\includegraphics[width=.4\textwidth]{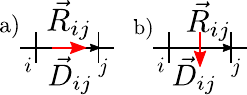}
\caption{The DMI vector $\vec{D}_{ij}$ pointing parallel a) or perpendicular 
b) to the vector $\vec{R}_{ij}$ between the $i$-th and $j$-th site
yielding {\color{black} Bloch- or N\'eel-type skyrmions-like structures} respectively.} 
\end{figure}
The magnitude $|\vec{D}_{ij}|$ is rather high for macroscopic systems containing
skyrmions with diameters up to $\sim 100\,{\rm nm}$, but in 
agreement with the choice from Ref. \onlinecite{PhysRevResearch.4.043113}
for skyrmions on quantum level.
The evolution of the magnetization 
pattern is controlled by the out-of-plane component $B_z$ of the external 
magnetic field expressed in the units of $\mu_B|J_\parallel |$, where $\mu_B$
is the Bohr magneton.  {\color{black} Open boundary conditions are taken
into account.}\par
A special consideration is given to the calculation of the topological charge
$Q$ defined by Eq. (1). The diagonalization of the Hamiltonian (2) yields
the magnetization $\vec{m}(x_i,y_i)$ at the mesh nodes with the coordinates
$(x_i,y_i)$, obtained as the Pauli-matrices
expectation values $\vec{m}(x_i,y_i)\equiv\left(\left<\sigma^x_i\right>,\left<
\sigma^y_i\right>,\left<\sigma_z^i\right>\right)$.
In order to calculate the partial derivatives and the surface
integral in (2) the field $\vec{m}(x,y)$ needs to be interpolated between
the mesh points.
The easiest way is to divide the mesh into triangles $t$, defined by three
neighboring nodes $t\equiv(i,j,k)$. In linear approximation the magnetization  
is a sum over all triangles $\vec{m}(x,y)\approx\sum_t
\vec{m}_t(x,y)$, where inside the triangle $t$: $\vec{m}_t(x,y)=\vec{a}_tx+
\vec{b}_ty+\vec{c}_t$ and outside the triangle $t$: $\vec{m}_t(x,y)=0$. 
The vectors $\vec{a}_t$, $\vec{b}_t$ and $\vec{c}_t$ are determined from
the magnetization values at the $i$, $j$ and $k$ nodes, obtained
as the expectation values of the corresponding Pauli matrices for the associated
qubits. It is easy to see that
the topological charge (1) in this approximation is a sum over all triangles
with the areas $S_t$: $q\approx(1/4\pi)\sum_t S_t(\vec{a}_t\times\vec{b}_t)\cdot
\vec{c}_t$. This approach of applying the concept of the topological charge
on discrete systems can be regarded as a simplified version of a more 
sophisticated winding number.\cite{BERG1981412,4121581,PhysRevX.9.041063}\par
{\color{black} The diagonalization of the Hamiltonian (2) 
is performed by applying the 
VQE method, implemented in the open-source Python package 
Tangelo\cite{tangelo}.}
We emulate the {\color{black} $N$-qubits} quantum circuit on the noiseless 
simulator backend, 
which is  incorporated in the package. The eigenstate $\left|\Phi\right>$ 
is expressed by 
the hardware-efficient ansatz (HEA)\cite{Kandala2017}, composed of an entangler
$U_{\rm ENT}$, inserted between two sets $i=1,2$ of the Euler rotations:
$\left|\Phi\right>=\prod_{q=1}^N\left[U(\vec{\theta}^{q,1})\right]
\times U_{\rm ENT}\times
\prod_{q=1}^N\left[U(\vec{\theta}^{q,2})\right]
\left|00\ldots 0\right>$, where $U(\vec{\theta}^{q,i})$
is expressed by the $R_x(\theta)$ and $R_x(\theta)$ gates
as 
$U(\vec{\theta}^{q,i})=R_z(\theta_1^{q,i})R_x(\theta_2^{q,i})
R_z(\theta_3^{q,i})$ The angles
$\vec{\theta}^{q,i}=(\theta_1^{q,i},\theta_2^{q,i},\theta_3^{q,i})$
represent $6\times N$ variational parameters that need to be optimized in 
order to find the $N$-qubits ground state associated with the energy $E$.\par
\begin{figure}
\includegraphics[width=.5\textwidth]{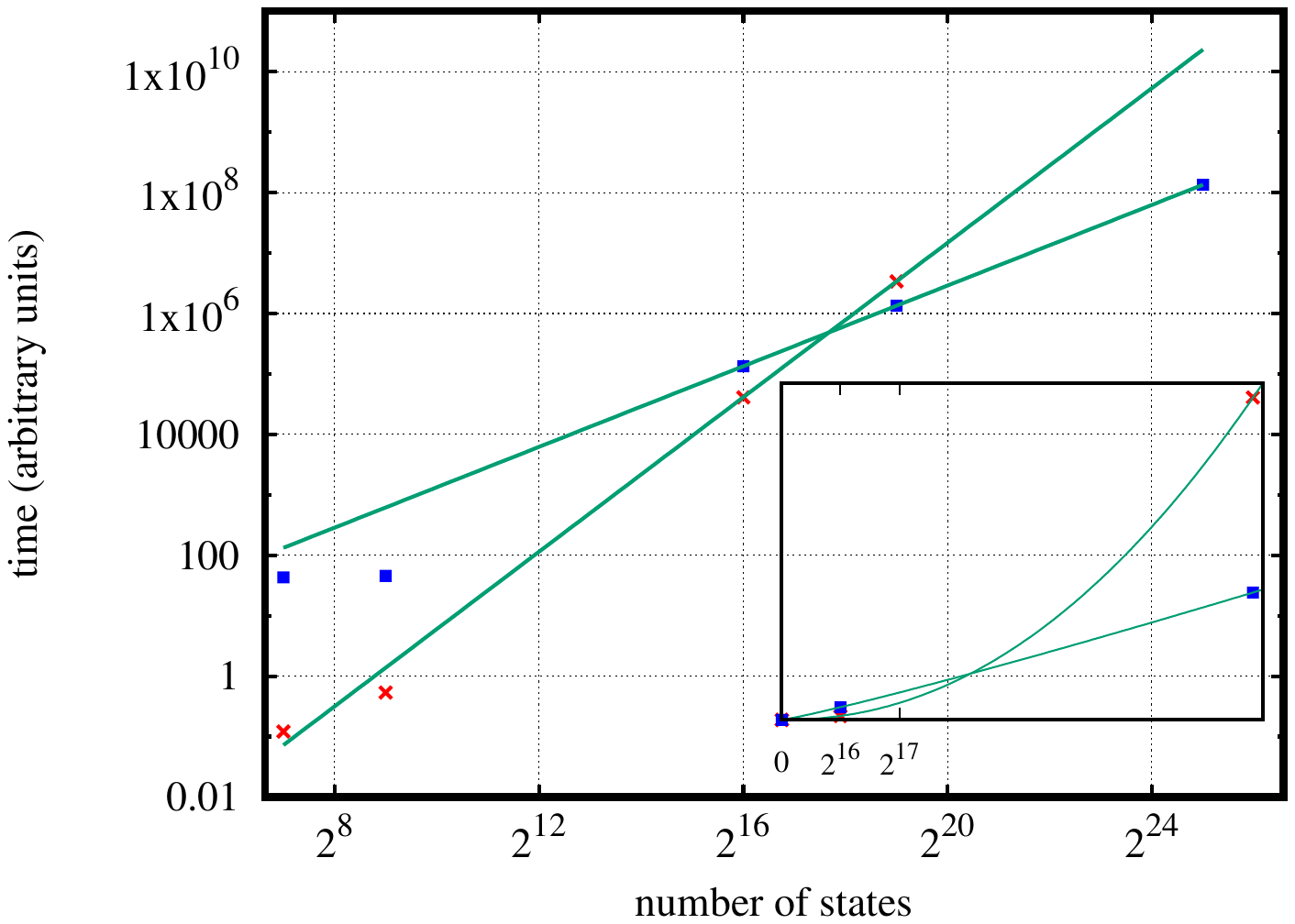}
\caption{\color{black}A comparison between in the performance of the
{\color{black} VQE
(${\color{blue}\blacksquare}$) and the Lanczos (${\color{red}\times}$)}
algorithms using a logarithmic scale. The scalabilities are ${\cal O}(n^{1.1})$
and ${\cal O}(n^{2.1})$ with the number $n$ of sites. The linear-scale
inset indicates the VQE being more efficient for $N>17$.}
\end{figure}
{\color{black} A power of the VQE algorithm is demonstrated by means of 
a comparison 
with the performance of the Implicitly Restarted Lanczos 
Method\cite{Arpack,Lehoucq1998} run serially on the same platform: FIG. 2. 
The tests are carried out on square lattices with $N=9,16$ and 25 
(only the VQE) sites, as well as, for a better statistics, on
triangular lattices with $N=7$ and 19. The corresponding fits reveal
that the VQE scales nearly linearly ${\cal O}(N^{1.1})$ and the 
Lanczos algorithm slightly worse than quadratically ${\cal O}(N^{2.1})$
for finding the ground state of the Hamiltonian (2). This comparison
clearly indicates the VQE as faster for $N>17$ which makes solving
the problem with the classical algorithm considerably more time consuming
already for a $5\times 5$ square lattice characterized by $2^{25}$ states.}
{\color{black} Furthermore, a symmetry breaking is known to slow the VQE 
convergence, which can be improved by applying a specially-prepared
ansatz\cite{PhysRevResearch.4.033100,Bertels2022}. This issue is not 
considered in the present work, where the only criterion is an agreement
with the Lanczos-method results within the numerical accuracy. Hence.
a further optimization might result in an even better VQE performance.
}
\par
{\color{black} The subsequent calculations are performed by applying the VQE
algorithm on a square lattice with $N=16$. We stick to the lattice size at
which the classical approach is still slightly faster because collecting
all the data points, presented in FIGs. 3 and 5 for $N>17$ would be time 
consuming
by means of quantum computing too, but not necessary for a proof of concept.
However, in the following we justify that it is completely achievable.}
The results for $\vec{D}_{ij}\parallel\vec{R}_{ij}$ and $J_\perp=
0.5|J_\parallel|$ are presented in FIG. 3. The energy $E$ linearly drops with
the field $B_z$ as expected from the Zeemnan interaction, described by the 
third term in Eq. (2), except for a transparent jump at $B_z=1.884$,
which clearly hints a transition in magnetic ordering. 
{\color{black} This is confirmed by
a change in the dependencies of the topological-charge $Q$ (1)
and of the magnetization $x$ component $m_x$ from a finite to 
a nearly-zero value.  The magnetization
components $m_\alpha$, $\alpha=x,y,z$ are calculated as the sums
of the expectation values of the corresponding Pauli matrices at
particular mesh nodes as: $m_\alpha=\sum_i\left<\sigma_i^\alpha\right>$.
\begin{figure}
\includegraphics[width=.5\textwidth]{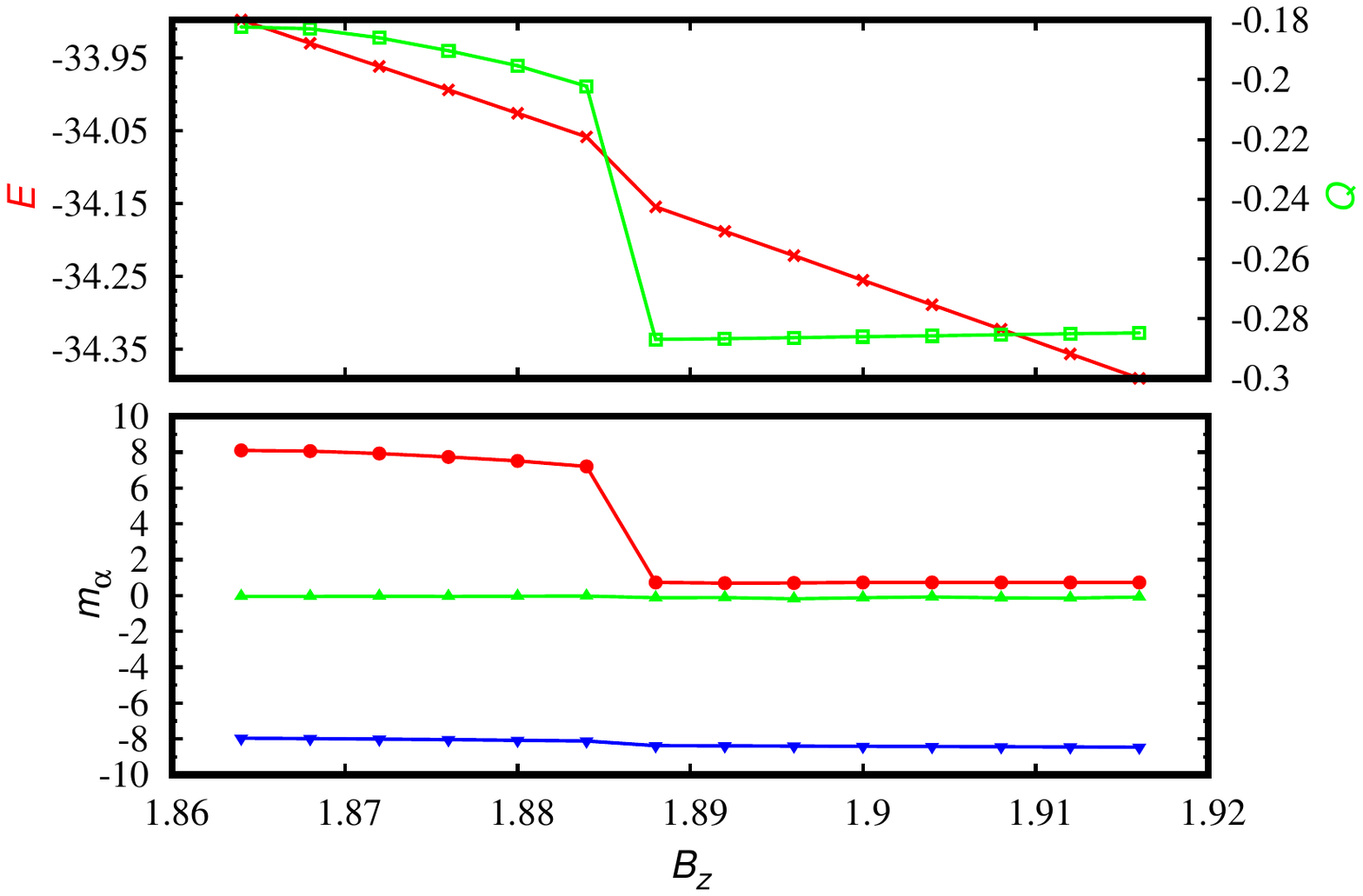}
\caption{\color{black} The external field $B_z$ dependence of the energy $E$ (${\color{red}
\times}$), the topological charge $Q$ (${\color{green}\square}$) and
the magnetization components $m_\alpha$ with $\alpha=x$ 
(${\color{red}\bullet}$) $y$ (${\color{green}\blacktriangle}$) and
$z$ (${\color{blue}\blacktriangledown}$) for $\vec{D}_{ij}
\parallel\vec{R}_{ij}$. The $B_z$ and $E$ are given in the dimensionless units
relative to the $|J_\parallel|$.}
\end{figure}
There is apparently no transition of the $m_y$ and $m_z$ components,
which remain nearly zero and finite, respectively.}
The assumption about the magnetization-pattern transition is further 
demonstrated in FIG. 4. Whereas there is a certain ordering at $B_z<1.884$,
{\color{black} reflected in a finite magnetization along the $x$ axis
($m_x$ in FIG. 3),}
a vortex-like spiral structure, resembling a Bloch-type {\color{black}
skyrmion-like structure}
with helicity $\gamma=\pi/2$, is formed for 
$B_z>1.884$, accompanied by a jump
in $|Q|$ by more than 50\% and a simultaneous drop in the {\color{black}
$m_x$, creating a certain $x$-$y$-plane symmetry in the magnetization pattern.}
\begin{figure}
\includegraphics[width=.4\textwidth]{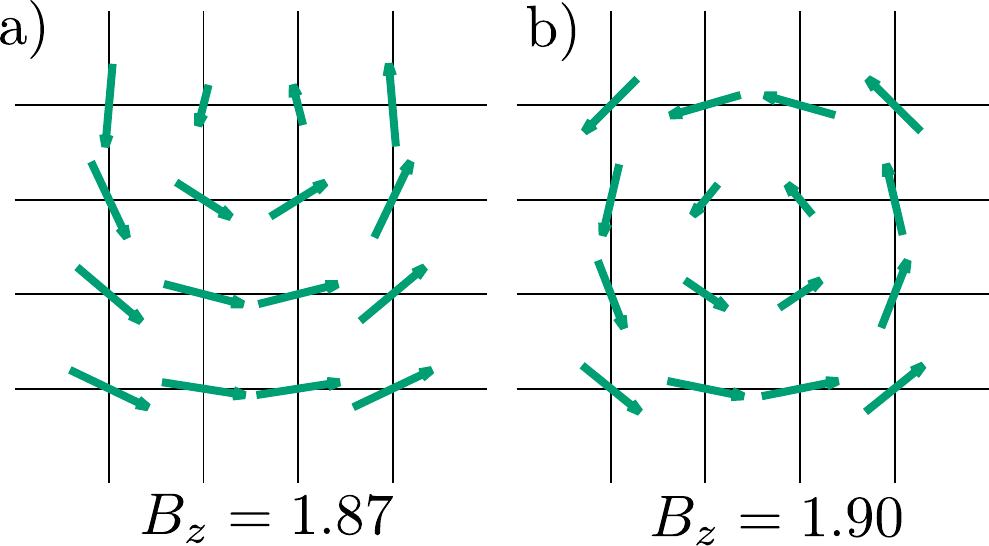}
\caption{The magnetization pattern for $\vec{D}_{ij}\parallel\vec{R}_{ij}$,
$J_\perp=0.5|J_\parallel|$
at $B_z=1.87$ a) and $B_z=1.90$ b).}
\end{figure}
\par
The situation is very similar in the case of $\vec{D}_{ij}\perp\vec{R}_{ij}$
and {color{black} $J_\perp=0.25\left| J_\parallel\right|$}: FIGs. 5,6. {\color{black} In this case, $J_\perp=
0.25\left| J_\parallel\right|$, used for $\vec{D}_{ij}\parallel\vec{R}_{ij}$, 
does not yield any transition to 
skyrmion-like ordering.}
\begin{figure}
\includegraphics[width=.5\textwidth]{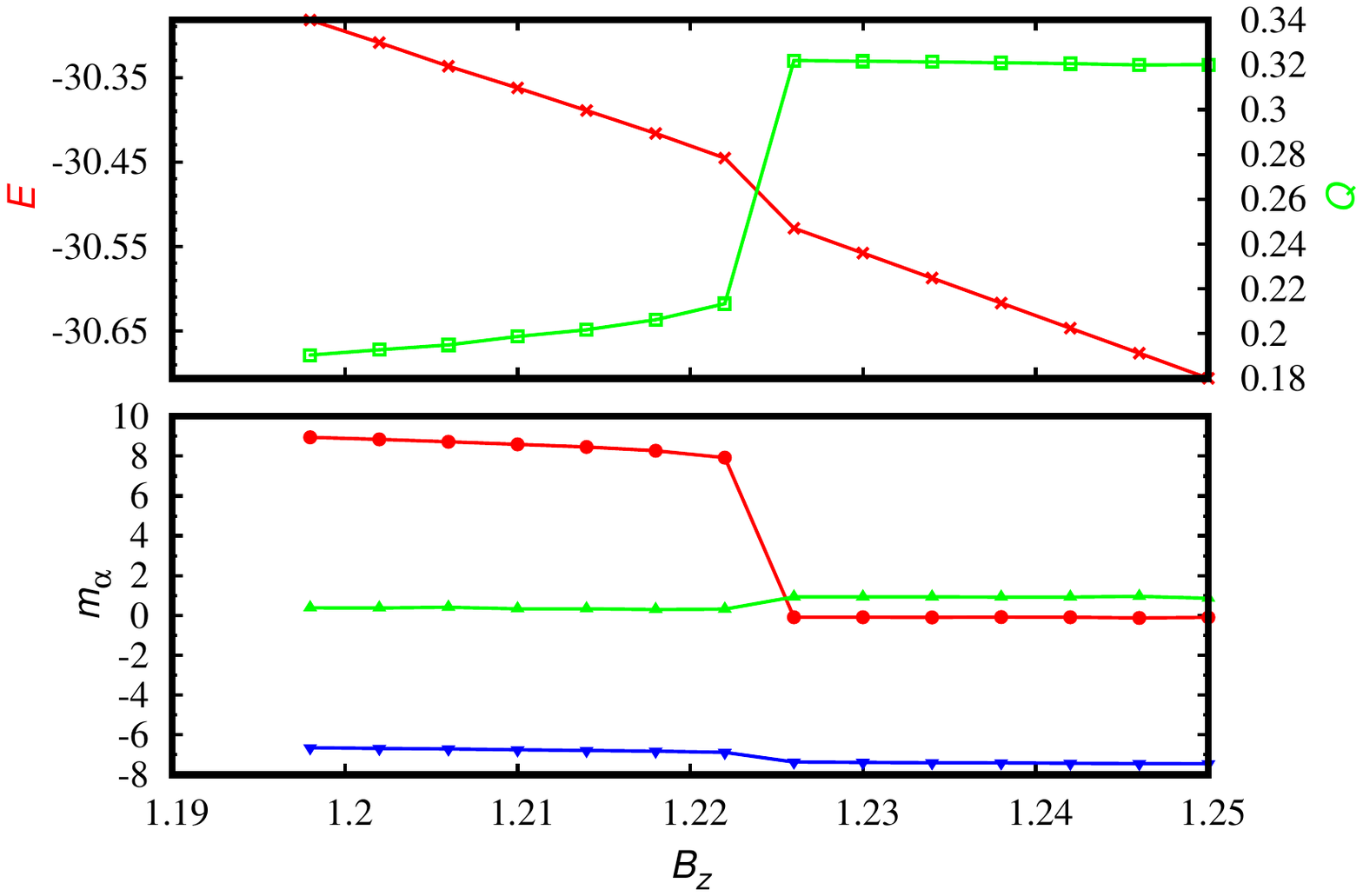}
\caption{\color{black} The external field $B_z$ dependence of the energy $E$ (${\color{red}
\times}$), the topological charge $Q$ (${\color{green}\square}$) and
the magnetization components $m_\alpha$ with $\alpha=x$ 
(${\color{red}\bullet}$) $y$ (${\color{green}\blacktriangle}$) and
$z$ (${\color{blue}\blacktriangledown}$) for $\vec{D}_{ij}
\perp\vec{R}_{ij}$. The $B_z$ and $E$ are given in the dimensionless units
relative to the $|J_\parallel|$.}
\end{figure}
\begin{figure}
\includegraphics[width=.4\textwidth]{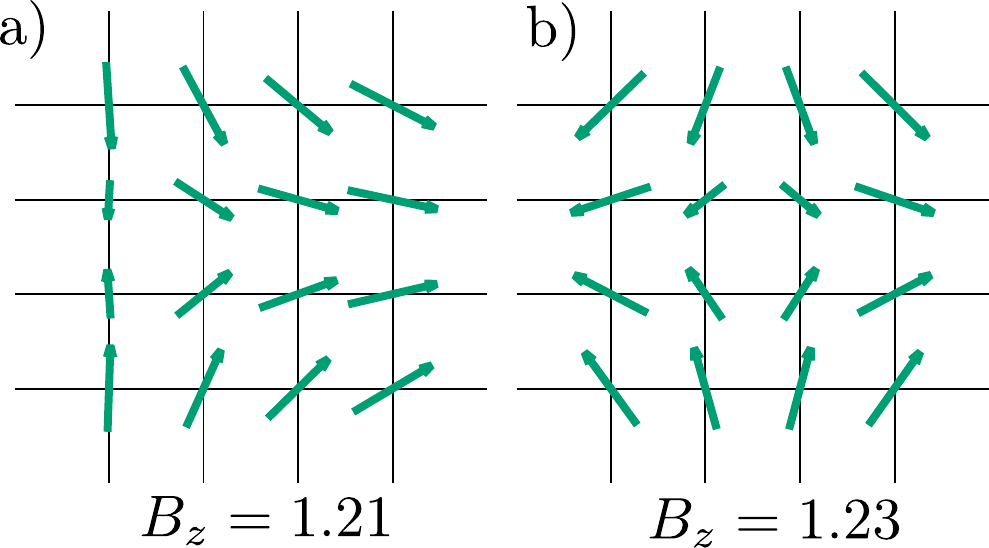}
\caption{The magnetization pattern for $\vec{D}_{ij}\perp\vec{R}_{ij}$,
$J_\perp=0.25|J_\parallel|$
at $B_z=1.21$ a) and $B_z=1.23$ b).}
\end{figure}
The choice of the exchange and the DMI coupling parameters is quite arbitrary
and it does not qualitatively influence the results. Contrary to the previous
case, the topological charge $Q$ is positive, hence, up to the transition point
at $B_z=1.222$ it parabolically increases, whereas the magnetization 
{\color{black} component $m_x$} drops. {\color{black} In this case, a small 
transition
is observable in the $m_y$ and $m_z$ components too. However, it is obvious
that an in-plane-symmetric state, similar to a N\' eel-type-skyrmion-like
structure with helicity $\gamma=0$,  is transformed from an 
$x$-direction-preferential ordering in FIG. 6 a). }
\par
Neither the pattern in FIG. 4 b), nor the pattern in FIG. 6 b) can be regarded
as proper skyrmions since they are not characterized by integer topological 
charges. The reason is probably a restricted geometry due to a limited number of
applied mesh nodes- qubits, which does not make a creation of a complete 
structure possible, {\color{black} although a VQE calculation on a $5\times 5$
square lattice for $\vec{D}_{ij}\parallel\vec{R}_{ij}$ reveals the existence of
a Bloch-type-skyrmion-like pattern at $B_z= 1.5$: FIG. 7. }
\begin{figure}
\includegraphics[width=.3\textwidth]{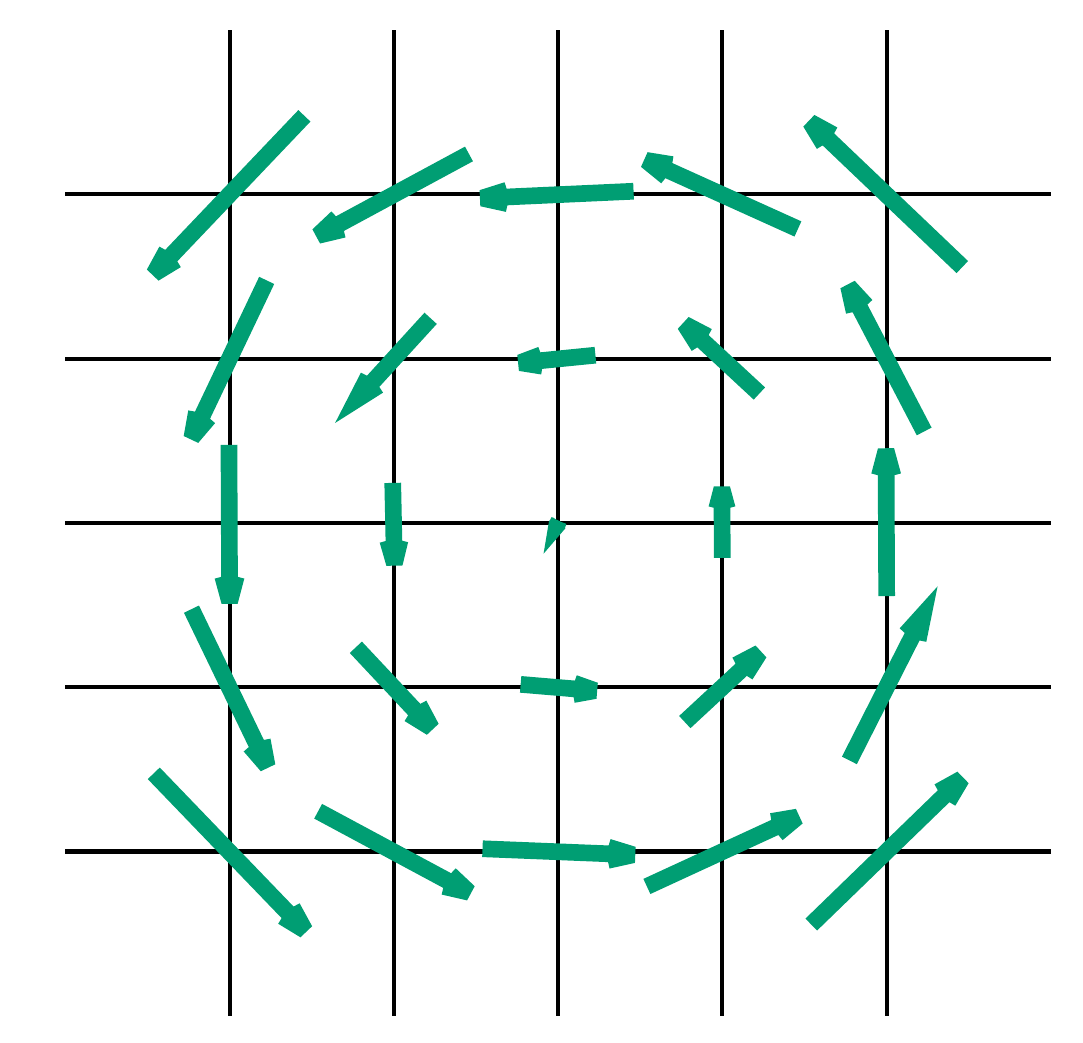}
\caption{\color{black} The magnetization pattern for 
$\vec{D}_{ij}\parallel\vec{R}_{ij}$, {\color{black} $J_\perp=0.5|J_\parallel|$} 
and $B_z=1.5$ on a $5\times 5$ square 
lattice.}
\end{figure}
{\color{black} Hence, it is very likely that}
the external-field $B_z$ driven transitions,
from partially disordered to magnetically ordered states, demonstrated in FIGs. 
3 and 5, clearly indicate a formation of new, skyrmionic-like phases in 
agreement with the assumption about the existence of quantum skyrmions at
$T=0$\cite{Istomin2000,doi:10.1142/9789811231711_0004,PhysRevResearch.4.043113}.
{\color{black} Although the energy difference of about 
$0.1\left|J_\parallel\right|$ between two phases at the transition point
is of the same order of magnitude as, for example, the 
magneto-crystalline-anisotropy energy, preventing a demagnetization
of a ferromagnet, it does not itself guarantee a stability of a skyrmion
as an information carrier. Nevertheless, it was argued that a similar energy
landscape yielded sufficiently long lifetimes for stable single
skyrmionic bits\cite{Hagemeister2015}.}
The present work demonstrates quantum computing by means of the VQE as
a suitable and a promising tool for investigations of quantum magnetism. The
results predict the existence of magnetic {\color{black} skyrmions-like 
structures} due to an anti-symmetric
DMI at the quantum level and call for eventual experiments on 
{\color{black} new}
materials in order to detect and characterize  such objects, {\color{black} to
apply them as potential carriers of 
information, and to implement the respective gates.}\cite{petrović2024colloquiumquantumpropertiesfunctionalities}.
\end{document}